\documentclass[onecolumn, floatfix, preprintnumbers, eqsecnum, letterpaper, superscriptaddress, nofootinbib]{revtex4}

\usepackage{graphicx}
\usepackage{microtype}
\usepackage{amsmath}
\usepackage{amssymb}
\usepackage{subfigure}
\usepackage{hyperref}
\usepackage{url}
\usepackage{xcolor}
\usepackage{color}
\usepackage{mathrsfs}
\usepackage{calrsfs}
\usepackage{amsfonts}
\usepackage{eufrak}
\usepackage{tabularx}
\usepackage{eucal}
\usepackage{latexsym}
\usepackage{ragged2e}
\usepackage{epsfig}
\usepackage{textcomp}

\usepackage{caption}
\usepackage{subfigure}

\usepackage{marvosym}
\usepackage{ifsym}

\usepackage{lipsum}

\DeclareCaptionJustification{justified}{\leftskip=0pt \rightskip=0pt \parfillskip=0pt plus 1fil}
\definecolor{vividviolet}{rgb}{0.62, 0.0, 1.0}
\definecolor{amaranth}{rgb}{0.9, 0.17, 0.31}
\definecolor{palatinateblue}{rgb}{0.15, 0.23, 0.89}
\definecolor{brightpink}{rgb}{1.0, 0.0, 0.5}
\definecolor{cornflowerblue}{rgb}{0.39, 0.58, 0.93}
\definecolor{deepcarminepink}{rgb}{0.94, 0.19, 0.22}
\definecolor{radicalred}{rgb}{1.0, 0.21, 0.37}

\hypersetup{ linktoc=all,
    colorlinks, linkcolor={palatinateblue},
    citecolor={brightpink}, urlcolor={amaranth}
}

\graphicspath{{Images/}}

\graphicspath{{Images/}}

\def\@fnsymbol#1{\ensuremath{\ifcase#1\or \ddagger \or  $\textleaf$  \or \dagger
\else\@ctrerr\fi}}%

\makeatother

\def\sideremark#1{\ifvmode\leavevmode\fi\vadjust{\vbox to0pt{\vss
 \hbox to 0pt{\hskip\hsize\hskip1em
 \vbox{\hsize1.3cm\tiny\raggedright\pretolerance10000
 \noindent #1\hfill}\hss}\vbox to8pt{\vfil}\vss}}}%

\def\beq{\begin{equation}}
\def\eeq{\end{equation}}

 \newcommand{\be}{\begin{equation}}
	\newcommand{\en}{\end{equation}}

\begin{document}

\title{Investigating Graviton Mass Effects on Black Hole Lensing in dRGT Massive Gravity}


\author{Haximjan Abdusattar \Letter}
\email{axim@ksu.edu.cn}
\affiliation{School of Physics and Electrical Engineering, Kashi University, Kashi 844000, Xinjiang, China}

\author{Yu-Xuan Han}
\email{hanyuxuan@ksu.edu.cn}
\affiliation{School of Physics and Electrical Engineering, Kashi University, Kashi 844000, Xinjiang, China}

\author{Abdujappar Rusul}
\email{aai-pulsar@qq.com}
\affiliation{School of Physics and Electrical Engineering, Kashi University, Kashi 844000, Xinjiang, China}

\author{Shi-Bei Kong \Letter}
\email{shibeikong@ecut.edu.cn}
\affiliation{School of Science, East China University of Technology, \\
Nanchang 330013, Jiangxi, China}

{\let\thefootnote\relax\footnotetext{\vspace*{0.1cm}$^{\text{\Letter}}$ Corresponding Author}}

\begin{abstract}

In this paper, we delve into the gravitational lensing and photon trajectories in the vicinity of non-asymptotically flat black hole within the framework of de Rham-Gabadadze-Tolley (dRGT) massive gravity, incorporating a non-zero graviton mass. We assume that both the observer and the light source are located at a finite distance from the lens object, and calculate the gravitational deflection angle of light ray by such a black hole based on Gauss-Bonnet theorem. Furthermore, we analytically derive the angular radius expression of Einstein rings in dRGT massive gravity and briefly discuss the relevant impacts of the graviton mass. Notably, our analytical results suggest potential impacts of the graviton mass on both light deflection angles and Einstein ring characteristics, highlighting its relevance in dRGT massive gravity and expecting the detectability of black holes in gravitational lensing observations.


\end{abstract}

\maketitle

\section{Introduction}\label{sect:In}

Astrophysical observations such as supernova of IA type \cite{Phillips:1993ng,SupernovaSearchTeam:1998fmf}, Cosmic Microwave Background radiation (CMB) \cite{Planck:2015fie,WMAP:2003elm} and baryon acoustic oscillations (BAO) \cite{Beutler:2011hx,SDSS:2009ocz} have indicated that the expansion of the universe is currently undergoing an acceleration phase. However, general relativity can not explain the origin of such phenomena satifactorily \cite{Weinberg:1988cp,Peebles:2002gy}. In Einstein's theory of general relativity, the graviton is a massless particle. Therefore, a natural and important question is whether the self-consistent theory of quantum gravity has massive gravitons. It is noticeable that the de Rham-Gabadadze-Tolley (dRGT) massive gravity theory can be considered as a modification of general relativity by providing consistent interaction terms which are interpreted as the graviton mass \cite{deRham:2010ik,deRham:2014zqa}.
This theory may provide a possible explanation for the current accelerated expanding of the universe and has received significant attention.

It is well known that black holes are the best objects for studying the modified theories of general relativity \cite{Psaltis:2008bb}. In this respect, some studies have been focused on finding the spherically symmetric black hole solutions in dRGT massive gravity \cite{Cai:2014znn,Xu:2015rfa,Ghosh:2015cva}.
Among them, the new exact solution obtained under this framework \cite{Ghosh:2015cva} has aroused growing interest in recent years. Note that the dRGT black hole solution can be regarded as the extension of solutions from Einstein field equations, such as the (Anti-)de Sitter-Schwarzschild solution associated with the graviton mass which naturally generates the cosmological constant. The thermodynamic properties of this black hole have already been discussed extensively \cite{Ghosh:2015cva,Chabab:2019mlu,Hou:2020yni,Nam:2020gud}, but on the contrary, its geometrical properties have rarely been studied. Therefore, many important geometric aspects should be thoroughly investigated for such a black hole, e.g. the gravitational bending and Einstein ring in the presence of the graviton mass.

To investigate the Einstein ring of the black hole, one usually calculates the gravitational deflection angle of the light ray. Indeed, light rays passing around the black hole have a large deviation, and even undergo many complete loops around the object before reaching the observer, resulting in an infinite set of relativistic images on each side of the object. By investigating the deflection angle, we could not only extract the information about black holes in the universe, but also verify profound alternative theories of gravity in their weak and strong field regimes from these relativistic images \cite{Bartelmann:1999yn,Schmidt:2008hc,Hoekstra:2013via}. In Ref.\cite{Gibbons:2008rj}, the authors apply the Gauss-Bonnet theorem (GBT) to derive the deflection angle of light in static asymptotically flat spacetime, where they assumed that the source and receiver are located at an infinite region. In later developments, a new step extends such investigation to more general situations \cite{Ishihara:2016vdc,Ishihara:2016sfv} where the receiver and source are assumed to be at finite distance from the lens.
Subsequently, the weak gravitational deflection of light ray by different lens objects have been widely studied \cite{Ovgun:2018tua,Ono:2019hkw,Jusufi:2018jof,Crisnejo:2018uyn,Pantig:2020odu,Fu:2021akc,Javed:2020fli,Ovgun:2018fnk,Jusufi:2017uhh,deLeon:2019qnp,Ono:2018jrv,Crisnejo:2019ril,Ovgun:2019wej,Li:2020zxi,He:2020eah,Pantig:2022toh,Qiao:2021trw}.
Both in the case of finite and infinite distance deflection, the Gauss-Bonnet theorem is often applied to an infinite region outside of the particle ray.

However, in the non-asymptotically flat spacetime, it can never assume that the source of light is located at infinite distance from a gravitational lens object. Due to this problem, Ref.\cite{Arakida:2017hrm} first considered a finite region and study the deflection angle of light in the Schwarzschild-dS spacetime. Moreover, by using the receiver, source, and the closest approach of the light ray, the deflection angle in non-asymptotically flat spacetime has also been studied \cite{Takizawa:2020egm} (See \cite{Haroon:2018ryd,Atamurotov:2021hoq,Gao:2023ltr} for more related works).
In \cite{Panpanich:2019mll}, the authors studied the gravitational deflection angle of black hole in dRGT massive gravity by using geodesic method, where they considered the special case that the distances from the gravitational lens to the source and observer are equal.
In the present work, we apply the Gauss-Bonnet theorem to investigate the weak field limit gravitational lensing of the black hole in dRGT massive gravity by considering a receiver and source at finite distance from a lens object.
Furthermore, we derive the analytic expression of the angular radius of Einstein ring for the black hole in dRGT massive gravity .

The outline of this paper is as follows. In section \ref{SSSGB}, we make a brief review on the calculation of the weak deflection angle of light rays for static spherically symmetric spacetime by Gauss-Bonnet theorem.
In section \ref{dRGT}, we derive the weak deflection angle of the light ray by an extended non-asymptotically flat spacetime that for the black hole in dRGT massive gravity, and demonstrate the influence of graviton mass on the radius of the photon sphere, impact parameter and deflection angle. In section \ref{observe}, we investigate the angular radius of the Einstein ring for the black hole in dRGT massive gravity and discuss the influence of the graviton mass on it.
In section \ref{summary}, we make conclusions and discuss possible future works.

\section{deflection angle of light rays for static spherically symmetric spacetime by Gauss-Bonnet theorem}\label{SSSGB}

In this section, we make a brief review of the weak deflection angle of light ray in a static spherically symmetric spacetime based on the Gauss-Bonnet theorem, in which receiver and source locate at finite distance from the lens object.

The line element in an isotropic coordinate system, $i.e.,$ ($t, r, \theta, \varphi$) for static spherically symmetric spacetime can be written as
\begin{equation}\label{SSSS}
{\rm d}s^2=-A(r){\rm d}t^2+B(r){\rm d}r^2+ C(r) ({\rm d}\theta^2+\sin^2\theta{\rm d}\varphi^2)\,,
\end{equation}
where $A(r)$, $B(r)$ and $C(r)$ are positive such that a static emitter and a static receiver can exist.

For the trajectory of the
light, we impose the null condition $ds^2 = 0$ \cite{Bozza:2002zj}. Thus, from Eq.(\ref{SSSS}) one can get
\begin{eqnarray}\label{gamma}
dt^2 &=& \gamma_{ij} dx^i dx^j \\
&=&\frac{B(r)}{A(r)} dr^2 + \frac{C(r)}{A(r)} d\varphi^2 \,,
\end{eqnarray}
where $i$, $j$ takes $1$, $2$, $3$, and $\gamma_{i j}$ is often called the optical metric (denoted as $M^{\mbox{opt}}$) \cite{Gibbons:2008rj,Gibbons:2008hb}, in which the light ray is a spatial geodesic curve.

Without loss of generality, we consider the equatorial plane $\theta=\pi/2$ to analyze the light trajectory. For a static, spherically symmetric spacetime, the impact parameter $b$ of light can be defined using its conserved energy $E$ and angular momentum $L$
~\cite{Ishihara:2016vdc}
\begin{align}\label{b}
b\equiv\frac{L}{E}=\dfrac{C(r)}{A(r)}\dfrac{d\varphi}{dt}\,.
\end{align}
From the above definition, and by setting $C(r)=r^2$, one further obtain \cite{Ishihara:2016vdc}
\begin{align}\label{orbiteq}
\left(\frac{du}{d\varphi}\right)^2=\frac{1}{b^2 A(u) B(u)}-\frac{u^2}{B(u)}\equiv F(u)\,,
\end{align}
where $u \equiv 1/r$ (with $r$ being the radial coordinate). For later convenience, we denote
$u_S$ and $u_R$ as the values of $u$ at the light source and receiver, respectively, and $u_0$ as the value at the closest point of the photon trajectory to the black hole.
For a finitely distant source and receiver, the light deflection angle $\hat{\alpha}$ is defined as~\cite{Ishihara:2016vdc,Ono:2019hkw}
\begin{eqnarray}
\hat{\alpha}\equiv\Psi_R-\Psi_S+\varphi_{RS},\label{definitionangle}
\end{eqnarray}
where $\varphi_{RS}$ represents the coordinate angle between the radial directions of the receiver and the source, given by
\begin{eqnarray}
\varphi_{RS}=\int_{u_S}^{u_0}\frac{du}{\sqrt{F(u)}}+\int_{u_0}^{u_R}\frac{du}{\sqrt{F(u)}},\label{phiRS}
\end{eqnarray}
and $\Psi_S$ ($\Psi_R$) is the angle between the light trajectory and the radial direction at the source (receiver).
Furthermore, following the notation in Ref.~\cite{Ishihara:2016vdc}, $\Psi$ denotes the angle between the tangent direction of the light trajectory and the radial coordinate at any point along the path. This angle satisfies
\begin{eqnarray}\label{sinPsi}
\sin\Psi=\dfrac{b\sqrt{A(r)}}{\sqrt{C(r)}}\,.
\end{eqnarray}
At present, the light deflection processing method in asymptotically flat spacetime is relatively mature and simple. However, it is difficult to calculate the deflection angle of light ray in non-asymptotically flat spacetimes. To solve this difficulty, Ref.\cite{Ishihara:2016vdc} proposed the method that apply the Gauss-Bonnet theorem to investigate the weak deflection angle of light rays by black holes. This method has been tested by large numbers of observations, and it has become widely adopted in physics and astronomy, when dueling with both weak and strong gravitational lensing.

The Gauss-Bonnet theorem establishes a fundamental connection between the intrinsic differential geometry and topology of a surface. Consider a compact, oriented, nonsingular two-dimensional Riemannian surface
$T$ with Euler characteristic $\chi(T)$ and Gaussian curvature $K$, whose boundary $\partial T$ is a piecewise smooth curve. The theorem is formulated as~\cite{Gibbons:2008rj,Ono:2019hkw,Qiao:2021trw}
\begin{align}\label{localGB}
\iint_{T}KdS+\oint_{\partial T}k_{g}dl+\sum\theta_a=2\pi\chi(T),
\end{align}
where $dS=\sqrt{\gamma_{rr}\gamma_{\varphi\varphi}}drd\varphi$ denotes the area element of the surface, $dl$ denotes the line element along the boundary, $\theta_a$ represents the external angle at the $a$-th vertex, $K$ is the Gaussian curvature of the optical space, expressed as \cite{Gibbons:2008rj,Ishihara:2016vdc}
\begin{eqnarray}\label{Gauss-K}
K=-\dfrac{1}{\sqrt{\gamma_{rr}\gamma_{\varphi\varphi}}}\left[\dfrac{\partial}{\partial r}\left(\dfrac{1}{\sqrt{\gamma_{rr}}}\dfrac{\partial\sqrt{\gamma_{\varphi\varphi}}}{\partial r}\right)+\dfrac{\partial}{\partial \varphi}\left(\dfrac{1}{\sqrt{\gamma_{\varphi\varphi}}}\dfrac{\partial\sqrt{\gamma_{rr}}}{\partial \varphi}\right)\right]\,
\end{eqnarray}
and $k_g$ is the geodesic curvature of a smooth curve $C$ defined by $r(\varphi)$. Taking $C_0=$ constant, $k_g$ is expressed~\cite{Gao:2023ltr,Li:2019vhp}
\begin{eqnarray}\label{geodesiccurvature}
k_g(C_0)=|\nabla_{\dot{C_0}}\dot{C_0}|=\Gamma^r_{\varphi\varphi}(\dot{C}_0^\varphi)^2\,,
\end{eqnarray}
with $\dot{C_0}$ denoting the tangent vector along $C_0$, and $\Gamma^r_{\varphi\varphi}$ the Christoffel symbol. The component $\dot{C}_0^\varphi$ is derived from the unit speed condition $\gamma_{\varphi\varphi}\dot{C}^\varphi_0\dot{C}^\varphi_0=1$.

To investigate the deflection angle of light rays in non-asymptotically flat spacetime, in Ref.\cite{Takizawa:2020egm} the authors considered a specific two regions $D_R$ and $D_S$ (See corresponding schematic diagram and more detailed analyses of the weak gravitation lensing in literature \cite{Takizawa:2020egm}).
In this manner, Eq.(\ref{localGB}) can be expressed separately for both regions. Subsequently, by combining this result with Eq. (\ref{localGB}) itself, the deflection angle in the Gauss-Bonnet theorem takes the form \cite{Takizawa:2020egm}
\begin{eqnarray}\label{finallyangle}
\hat{\alpha}=\iint_{D_R+D_S}KdS+\int^{P_S}_{P_R}k_{g}(C_0)dl+\varphi_{RS}\,.
\end{eqnarray}

In the next section, we apply Eq.(\ref{finallyangle}) to calculate the weak light deflection angle for a new class of dRGT massive gravity black holes.

\section{deflection angle of light ray by a new class of black hole in dRGT massive gravity}\label{dRGT}

In this section, we follow the procedure \cite{Ishihara:2016vdc,Takizawa:2020egm,Gao:2023ltr} and present the gravitational deflection angle of in new class of dRGT massive gravity black hole through the Gauss-Bonnet theorem.

The line element of the black hole in dRGT massive gravity is given by \cite{Ghosh:2015cva}
\begin{equation}\label{spacetime}
ds^2=-A(r)dt^2+B(r)dr^2+C(r)(d\theta^2+\sin^2\theta d\varphi^2)
\end{equation}
with\footnote{In the case $m_g\rightarrow0$, the black hole solution reduces to the Reissner-Nordstr\"{o}m (RN) black hole and Schwarzschild black hole solution while both of $m_g\rightarrow0$, $Q\rightarrow0$, as expected.
The solution can be classified according to the values of $\alpha$ and $\beta$. When $(1+\alpha +\beta) < 0$, the solution takes the form of a de Sitter (dS) spacetime; in contrast, when $(1+\alpha +\beta) > 0$, the solution corresponds to an Anti-de Sitter (AdS) spacetime \cite{Ghosh:2015cva}. In particular, when $\alpha=-1$ and $\beta=1/3$ $(\gamma=0, \zeta=0)$, one can see that the result recover a similar of RN AdS black hole where $-m_g^2$ play the role of the cosmological constant, $i.e.,$ $\Lambda=-m_g^2$ \cite{Chabab:2019mlu}.}
\begin{equation}\label{BH}
A(r)=B(r)^{-1}=f(r)\,,~~~~~f(r)=1-\frac{2 M}{r}+\frac{Q^2}{r^2}-\frac{\Lambda}{3}r^2+\gamma r+\zeta\,,\,\,\,\,\, C(r)=r^2 \,,
\end{equation}
where $M$ is the black hole mass parameter, and $\Lambda$, $\gamma$ and $\zeta$ are defined as
\begin{eqnarray}\label{CondI}
\Lambda=-3m_g^{2}\left(1+\alpha +\beta \right) \,,~~~~~~~~~~~
\gamma=-c m_g^{2}\left(1+2\alpha +3\beta \right) \,,~~~~~~~~~~~
\zeta=c^{2} m_g^{2}\left(\alpha +3\beta \right)\,,
\end{eqnarray}
all of which originate from the graviton mass $m_g$.
Note that the cosmological constant $\Lambda$ is $[L]^{-2}$; the graviton mass $m_g$ and the coupling parameter $\gamma$ are $[L]^{-1}$; the coupling constants $\alpha$, $\beta$ and the curvature parameter $\zeta$ are all dimensionless quantities $[L]^0$;
the dimension of $c$ is $[L]$, which is set to be $c=b$ in the following for the continuity of discussion.

The equation of the photon sphere is given by \cite{Bozza:2002zj}
\begin{equation}\label{photon sphere}
  \frac{C'(r)}{C(r)}=\frac{A'(r)}{A(r)}\,,
\end{equation}
which admits at least one positive solution and the largest real root of Eq.(\ref{photon sphere}) is defined as the radius of the photon sphere $r_{ph}$. Substituting (\ref{BH}) into (\ref{photon sphere}), we obtain the radius equation of photon sphere for the black hole in the dRGT massive gravity
\begin{equation}\label{Rph}
  \frac{3 \left\{r_{ph}[r_{ph} (2 \zeta +\gamma r_{ph}+2)-6 M]+4 Q^2\right\}}{\Lambda r_{ph}^5-3 r_{ph} \left\{r_{ph}[r_{ph} (\zeta +\gamma r_{ph}+1)-2 M]+Q^2\right\}}=0\,,
\end{equation}
which is consistent with the result in Ref. \cite{Panpanich:2019mll} and correctly yields $r_{ph}^{RN}$ in the RN metric limit $(i.e., m_g=0)$ \cite{Claudel:2000yi,Eiroa:2002mk} and in the Schwarzschild metric limit $(i.e., Q=0, m_g=0)$ \cite{Bozza:2002zj}.
It can be seen that the photon sphere equation does not depend on the cosmological constant. The radius of photon sphere is then determined from the roots of this equation. Since this is a cubic equation, it is possible to obtain three real roots. We should only take the largest one, which locates outside the event horizon.

Use Eqs.~\eqref{orbiteq} and (\ref{BH}), we obtain the orbit equation of light for the black hole in dRGT massive gravity given by
\begin{eqnarray}\label{dup}
\left(\frac{du}{d\varphi}\right)^2&=&\frac{1}{b^2}-u^2+2Mu^3 -Q^2 u^4+\frac{\Lambda}{3}-\gamma u -\zeta u^2\,,
\end{eqnarray}
where $u(\varphi)=1/r$. Due to the complexity of Eq.(\ref{dup}), its exact solution is difficult to derive. We therefore adopt the weak-field limit approximation in the subsequent analysis, so dimensionless quantity $M/b\ll 1$ is treated as a small quantity and we only retain its
first and second-order terms in our calculations. Recall that for the Reissner-Nordstr\"{o}m (RN) black hole to possess an event horizon~\cite{Eiroa:2002mk}, the charge $Q$ must not exceed the mass $M$, i.e. $Q^2/b^2 \ll M/b$. Additionally, within the framework of this work, we assume that the graviton mass exerts a weaker influence on spacetime geometry
than the black hole mass, namely $m_g b \ll M/b$, so $m_g b$ is also a small quantity and terms higher than three orders will be dropped.
To streamline the discussion, we introduce a dimensionless small parameter $\varepsilon$ and further stipulate $\mathcal{O}(m_g b)\sim \mathcal{O}(Q^2/b^2)\sim \mathcal{O}(M^2/b^2)\sim \mathcal{O}(\varepsilon^2)\ll 1$. All subsequent calculations will be carried out up to the second-order precision $\mathcal{O}(\varepsilon^2)$.

Then, considering the boundary condition $\big(\frac{du}{d\varphi}\big)_{\varphi={\pi/2}}=0$, we can get the solution of above equation by iterative method in the following
\begin{eqnarray}\label{Weylorbit}
u(\varphi)&=&\frac{\sin\varphi}{b}+\frac{M(1+\cos^2\varphi)}{b^2}+\frac{M^2[37 \sin\varphi-3 \sin3 \varphi+30 (\pi -2 \varphi ) \cos \varphi]}{16 b^3}-\frac{Q^2[6(\pi-2\varphi)\cos\varphi+9\sin\varphi+\sin 3\varphi]}{16 b^3} \nonumber\\
&&+\frac{b\Lambda \sin\varphi}{6}-\frac{\gamma}{2}-\frac{(\pi-2\varphi)\cos\varphi+2\sin\varphi}{4b}\zeta+\mathcal{O}(\varepsilon^3,...)\,.
\end{eqnarray}
Note that, if $\varphi=\pi/2$, we have $u(\pi/2)=u_0=1/r_0$. Then, from the above equation one can get
\begin{eqnarray}\label{Weylorbit0}
u(\pi/2)=u_0=\frac{1}{b}+\frac{M}{b^2}+\frac{5M^2}{2b^3}-\frac{Q^2}{2b^3}+\frac{b \Lambda}{6}-\frac{\gamma}{2}-\frac{\zeta}{2b}+\mathcal{O}(\varepsilon^3,...)\,.
\end{eqnarray}
In the present paper, we consider the distance from the source to the receiver is finite because every observed stars and galaxies are located at finite distance from us (e.g., at finite red-shift in cosmology) and the distance is much larger than the size of the lens.
Therefore, let $u_R$ and $u_S$ denote the inverse of $r_R$ and $r_S$, respectively, where $r_R$ and $r_S$ are finite. Straightforwardly, from (\ref{Weylorbit}), we can obtain two solutions of the coordinate angle as
\begin{eqnarray}\label{phiRS}
\varphi&=&\arcsin (b u)+\frac{M \left(b^2 u^2-2\right)}{b \sqrt{1-b^2 u^2}}+\frac{M^2 \Big(15 \arcsin(b u)-\frac{b u (3 b^4 u^4-20 b^2 u^2+15)}{\left(1-b^2 u^2\right)^{3/2}}\Big)}{4 b^2}-\frac{Q^2 \Big(b^3 u^3+3 \sqrt{1-b^2 u^2} \arcsin(b u)-3 b u\Big)}{4 b^2 \sqrt{1-b^2 u^2}}\nonumber\\
&&+\frac{\Lambda b u}{6\sqrt{1-b^2 u^2}}
+\frac{b\gamma}{\sqrt{1-b^2 u^2}} +\frac{b\zeta u}{\sqrt{1-b^2 u^2}} +\mathcal{O}(\varepsilon^3,...)\,.
\end{eqnarray}
In addition, the corresponding non-vanishing components of the optical metric (\ref{BH}) are
\begin{eqnarray}
\gamma_{r r}&=&\frac{B(r)}{A(r)}=\Big(1-\frac{2 M}{r}+\frac{Q^2}{r^2}-\frac{\Lambda}{3}r^2+\gamma r+\zeta \Big)^{-2}\,, \label{gammar} \\ \gamma_{\varphi\varphi}&=&\frac{C(r)}{A(r)}=\Big(1-\frac{2M}{r}+\frac{Q^2}{r^2}-\frac{\Lambda}{3}r^2+\gamma r+\zeta \Big)^{-1} r^2 \,. \label{gammaV}
\end{eqnarray}
Substituting (\ref{gammar}) and (\ref{gammaV}) into (\ref{Gauss-K}) the Gaussian curvature is given by
\begin{eqnarray}\label{KottlerGauss}
\mathcal{K}=-\frac{2M}{r^3}+\frac{3M^2}{r^4}+\frac{3Q^2}{r^4}
+\left(\frac{2M}{r}-\frac{1}{3}\right)\Lambda-\frac{3M}{r^2}\gamma-\frac{\gamma^2}{4}
-\left(\frac{2M}{r^3}+\frac{\Lambda}{3}\right)\zeta +\mathcal{O}(\varepsilon^3 ,...)\,.
\end{eqnarray}

Thus, we have
\begin{eqnarray}\label{TRS}
&&\iint_{T_R+T_S}\mathcal{K}dS=\iint_{T_R+T_S} \mathcal{K}\sqrt{\gamma_{r r}\gamma_{\varphi \varphi}}~dr d\varphi \nonumber\\
&=&\int_{\varphi_S}^{\varphi_R}\int_{u_0}^{u(\varphi)}\Big[2 M+3 M^2 u-3 Q^2 u+\frac{\Lambda ^2}{6 u^5}-\frac{\gamma  \Lambda }{2 u^4}+\frac{\gamma ^2}{4 u^3}-\frac{\zeta \Lambda }{6 u^3}+\frac{\Lambda }{3 u^3}\Big] ~du d\varphi \nonumber\\
&=&\int_{\varphi_S}^{\varphi_R} \Big[2M(u-u_0)+\frac{3}{2}M^2 (u^2-u_0^2)-\frac{3}{2}Q^2 (u^2-u_0^2)-\frac{\Lambda^2}{24}\Big(\frac{1}{u^4}-\frac{1}{u_0^4}\Big)+\frac{\gamma \Lambda}{6}\Big(\frac{1}{u^3}-\frac{1}{u_0^3}\Big)\nonumber\\
&&-\frac{\gamma^2}{8}\Big(\frac{1}{u^2}-\frac{1}{u_0^2}\Big)-\frac{(u^2 -u_0^2)(\zeta-2)\Lambda}{12u^2 u_0^2}\Big]~ d\varphi \nonumber\\
&=&\int_{\varphi_S}^{\varphi_R} \Big[\frac{2 M (\sin\varphi-1)}{b}+ \frac{2M^2 \cos^2\varphi}{b^2} +\frac{3Q^2 \cos^2\varphi}{2b^2}-\frac{b^2 \Lambda \cot^2 \varphi}{6} -\frac{b^4 \Lambda^2 }{144}(\cos 2\varphi+5) \cot^2 \varphi \csc^2\varphi-\frac{b^2 \gamma ^2 }{8} \cot^2 \varphi \Big]~ d\varphi \nonumber\\
&=& \frac{2M (\sqrt{1-b^2 u_R^2}+\sqrt{1-b^2 u_S^2})}{b}-\frac{3Q^2 (u_R\sqrt{1-b^2 u_R^2}+u_S\sqrt{1-b^2 u_S^2})}{4b}+\frac{M^2}{b^2}\left[\frac{b u_R(3-b^2 u_R^2)}{\sqrt{1-b^2 u_R^2}}+\frac{b u_S(3-b^2 u_S^2)}{\sqrt{1-b^2 u_S^2}}\right]\nonumber\\
&&-bM\gamma\Big(\frac{u_R}{\sqrt{1-b^2 u_R^2}}+\frac{u_S}{\sqrt{1-b^2 u_S^2}}\Big)
-bM\zeta\Big(\frac{u_R^2}{\sqrt{1-b^2 u_R^2}}+\frac{u_S^2}{\sqrt{1-b^2 u_S^2}}\Big) \nonumber\\
&&-\frac{b\Lambda}{6}\Big(\frac{\sqrt{1-b^2 u_R^2}}{u_R}+\frac{\sqrt{1-b^2 u_S^2}}{u_S}\Big)+\frac{b\gamma\Lambda}{12}\Big(\frac{1}{u_R^2\sqrt{1-b^2 u_R^2}}+\frac{1}{u_S^2\sqrt{1-b^2 u_S^2}}\Big)\nonumber\\
&&+\frac{\zeta\Lambda}{24}\Big[\Big(\frac{2b}{u_R \sqrt{1-b^2 u_R^2}}+\frac{2b}{u_S \sqrt{1-b^2 u_S^2}}\Big)-\Big(\frac{\pi}{u_R^2}+\frac{\pi}{u_S^2}\Big)\Big]
+\frac{b\Lambda^2}{72}\Big(\frac{b^2 u_R^2-2b^4 u_R^4-1}{u_R^3 \sqrt{1-b^2 u_R^2}}+\frac{b^2 u_S^2-2b^4 u_S^4-1}{u_S^3 \sqrt{1-b^2 u_S^2}}\Big)\nonumber\\
&&+\Big(-\frac{1}{72} b^4 \Lambda^2 +\frac{1}{8} b^2 \gamma^2 +\frac{1}{12} b^2 \zeta \Lambda +\frac{1}{6} b^2 \Lambda +\frac{M^2}{b^2}+\frac{3 Q^2}{4 b^2}-\frac{2 M}{b} \Big)\varphi_{RS} +\mathcal{O}(\varepsilon^3 ,...)\,,
\end{eqnarray}
where we used Eq.(\ref{Weylorbit}).
From (\ref{geodesiccurvature}), the geodesic curvature
\begin{eqnarray}\label{kgl}
\kappa_g(C_0)=-\frac{1}{r_0}+\frac{3M}{r_0^2}-\frac{2Q^2}{r_0^3}-\frac{\gamma}{2}-\frac{\zeta}{r_0} +\mathcal{O}(\varepsilon^3 ,...)\,,
\end{eqnarray}
and its integral are calculated as
\begin{eqnarray}\label{alpha-kgl}
G_c&=&\int_{P_R}^{P_S} \kappa_g(C_0)dl=\int_{\varphi_R}^{\varphi_S} \kappa_g(C_0)\frac{dl}{d\varphi}d\varphi \\
&=&\int_{\varphi_R}^{\varphi_S} \Big(\frac{\zeta^2}{8}-\frac{\zeta}{2}-\frac{M (2 \zeta +\gamma {r_0})}{2 {r_0}}+\frac{2 M}{{r_0}}-\frac{3 Q^2}{2 {r_0}^2}-\frac{\gamma^2 {r_0}^2}{8}+\frac{1}{12} (\zeta -2) \Lambda {r_0}^2-1\Big) d\varphi
\nonumber\\
&=&\int_{\varphi_R}^{\varphi_S} \Big(\frac{b^4 \Lambda ^2}{18}-\frac{b^3 \gamma \Lambda}{6} -\frac{b^2 \gamma^2}{8}-\frac{b^2 \zeta \Lambda}{12} -\frac{b^2 \Lambda}{6}+\frac{2 M^2}{b^2}-\frac{3 q}{2 b^2}-\frac{2 \zeta M}{b}+\frac{2 b \Lambda M}{3}+\frac{2 M}{b}+\frac{\zeta^2}{8}-\frac{\zeta }{2}-\frac{3 \gamma M}{2}-1 \Big) d\varphi\nonumber\\
&=&\Big[\frac{b^4 \Lambda^2}{18}-\frac{b^2 \gamma^2}{8}+\frac{2 M^2}{b^2}-\frac{3 Q^2}{2 b^2}-\frac{b \Lambda}{12}\Big(b (2 b \gamma +\zeta +2)-8 M\Big)-M \Big(\frac{2 \zeta }{b}+\frac{3 \gamma}{2}\Big)+\frac{2 M}{b}+\frac{\zeta}{8} (\zeta -4) \Big]\varphi_{RS}-\varphi_{RS} +\mathcal{O}(\varepsilon^3 ,...)\,.\nonumber
\end{eqnarray}

Finally, substituting Eqs.(\ref{phiRS}), (\ref{TRS}) and (\ref{alpha-kgl}) into (\ref{finallyangle}), the deflection angle becomes
\begin{eqnarray}\label{dRGTa}
\hat{\alpha}&=&\frac{2M (\sqrt{1-b^2 u_R^2}+\sqrt{1-b^2 u_S^2})}{b}-\frac{3Q^2 (u_R\sqrt{1-b^2 u_R^2}+u_S\sqrt{1-b^2 u_S^2})}{4b}+\frac{M^2}{b^2}\Big[\frac{b u_R(3-b^2 u_R^2)}{\sqrt{1-b^2 u_R^2}}+\frac{b u_S(3-b^2 u_S^2)}{\sqrt{1-b^2 u_S^2}}\Big]\nonumber\\
&&-\frac{b \gamma^2}{8} \Big(\frac{\sqrt{1-b^2 u_S^2}}{u_S}+ \frac{\sqrt{1-b^2 u_R^2}}{u_R} \Big)-\frac{b \Lambda }{6} \Big(\frac{\sqrt{1-b^2 u_R^2}}{u_R}+\frac{\sqrt{1-b^2 u_S^2}}{u_S}\Big)-b \zeta  M \Big(\frac{u_R^2}{\sqrt{1-b^2 u_R^2}}+\frac{u_S^2}{\sqrt{1-b^2 u_S^2}}\Big)\nonumber\\
&&+\frac{b \gamma \Lambda}{12} \Big(\frac{1}{u_R^2 \sqrt{1-b^2 u_R^2}}+\frac{1}{u_S^2 \sqrt{1-b^2 u_S^2}}\Big)-\frac{b^2 \zeta  \Lambda}{24} \Big[\Big(\frac{\pi}{b^2 u_R^2}+\frac{\pi}{b^2 u_S^2}\Big)-\Big(\frac{2}{b u_R \sqrt{1-b^2 u_R^2}}+\frac{2}{b u_S \sqrt{1-b^2 u_S^2}}\Big)\Big]\nonumber\\
&&+\frac{b \Lambda^2}{72} \Big(\frac{-2 b^4 u_R^4+b^2 u_R^2-1}{u_R^3 \sqrt{1-b^2 u_R^2}}+\frac{-2 b^4 u_S^4+b^2 u_S^2-1}{u_S^3 \sqrt{1-b^2 u_S^2}}\Big)+\frac{\zeta^2}{4}\Big(\frac{b u_R}{\sqrt{1-b^2 u_R^2}} +\frac{b u_S}{\sqrt{1-b^2 u_S^2}}\Big) \nonumber\\
&&+\frac{b \zeta M}{2} \Big(\frac{ u_R^2}{\sqrt{1-b^2 u_R^2}}+\frac{ u_S^2}{\sqrt{1-b^2 u_S^2}}\Big)-\frac{\zeta M}{b} \Big(\frac{1}{\sqrt{1-b^2 u_R^2}}+\frac{1}{\sqrt{1-b^2 u_S^2}} \Big) +\frac{b \gamma \zeta}{4} \Big(\frac{1}{\sqrt{1-b^2 u_R^2}}+\frac{1}{\sqrt{1-b^2 u_S^2}}\Big)
\nonumber\\
&&+\frac{[\arccos(b u_R)+\arccos(b u_S)]}{24b^2}\Big[72 M^2-18 Q^2+b^6 \Lambda ^2-4 b^5 \gamma \Lambda +3 b^2 (\zeta -4) \zeta +4 b M (4 b^2 \Lambda -9 b \gamma -12 \zeta)\Big]\nonumber\\
&& -\frac{b^3 \Lambda \zeta}{12} \Big(\frac{u_R}{\sqrt{1-b^2 u_R^2}}+\frac{u_S}{\sqrt{1-b^2 u_S^2}}\Big) +\mathcal{O}(\varepsilon^3 ,...)\,,
\end{eqnarray}
which is the gravitational deflection angle of light traveling along null geodesics for black hole in dRGT massive gravity in cases where observer and light source are both at finite distance regions.\footnote{
Setting $m_g = 0$ recovers the Reissner-Nordstr\"om black hole \cite{Ishihara:2016vdc,Crisnejo:2019xtp}; $\gamma=\zeta = 0$ gives the form of Reissner-Nordstr\"{o}m-AdS result; and $\gamma=\zeta=Q = 0$ reduces to the form that of the Schwarzschild-AdS black hole \cite{Ono:2019hkw,He:2020eah}.}
The electric charge clearly decreases the deflection angle, while the other modification terms is more complicated since it depends not only on the graviton mass, but also on the impact parameter and the black hole mass.

As can be readily seen from Eq.(\ref{dRGTa}), the deflection angle diverges as
$u_R\rightarrow0$ and $u_S\rightarrow0$, which fully reflects the non-asymptotically flat spacetime characteristic of this black hole. However, if we set $\Lambda=0$ and $\gamma=0$ first, and then consider the weak field approximation with $u_R\rightarrow0$, $u_S\rightarrow0$, we can rewrite the deflection angle of light as following
\begin{equation}\label{alpha1}
\hat{\alpha}=\frac{4M}{b}+\frac{3\pi M^2}{b^2}-\frac{3\pi Q^2}{4b^2} -\frac{\zeta [\pi b+4(1+\pi) M]}{2 b} +\frac{\pi \zeta ^2}{8}+\mathcal{O}(\varepsilon^3 ,...)\,.
\end{equation}

\begin{figure}[h]
\centering
\begin{minipage}[t]{7cm}
\centering
\includegraphics[width=6.5cm,height =5.2cm]{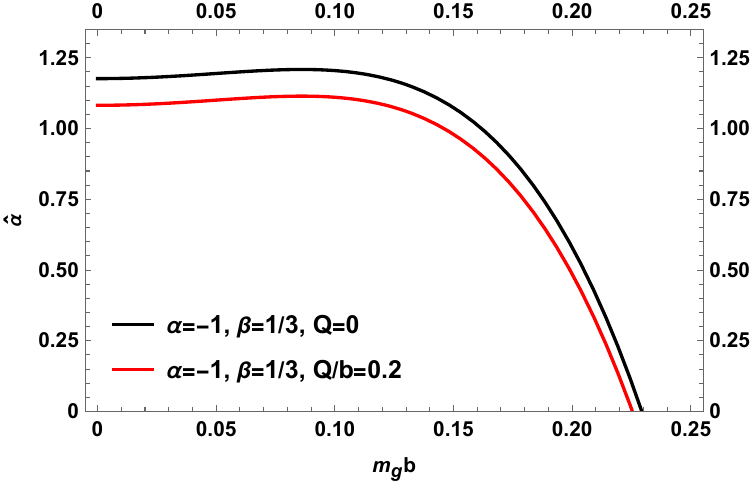}
\put(-100,-10){(a)}
\end{minipage}
\begin{minipage}[t]{7cm}
\centering
\includegraphics[width=6.5cm,height =5.2cm]{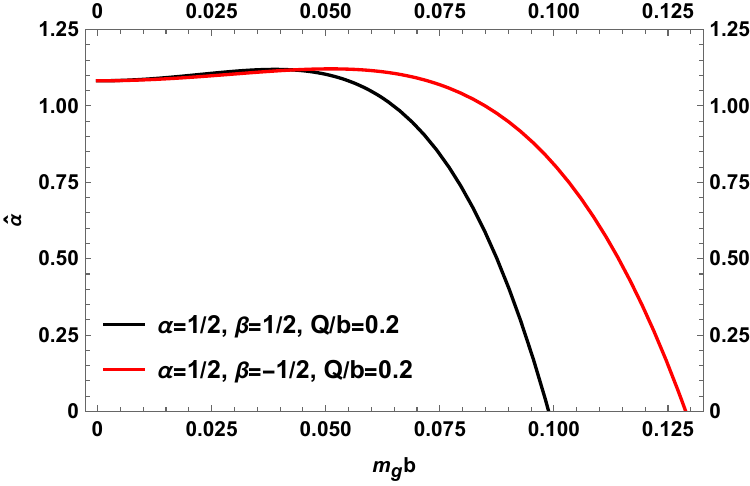}
\put(-100,-10){(b)}
\end{minipage}
\begin{minipage}[t]{7cm}
\centering
\includegraphics[width=6.5cm,height =5.2cm]{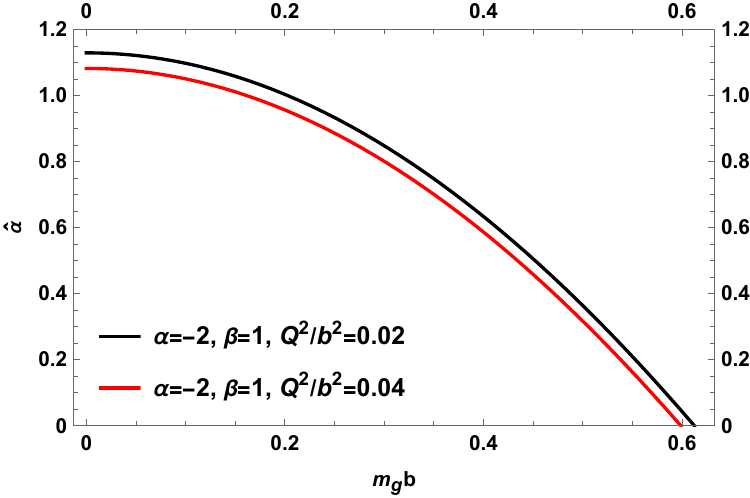}
\put(-100,-10){(c)}
\end{minipage}
\begin{minipage}[t]{7cm}
\centering
\includegraphics[width=6.5cm,height =5.2cm]{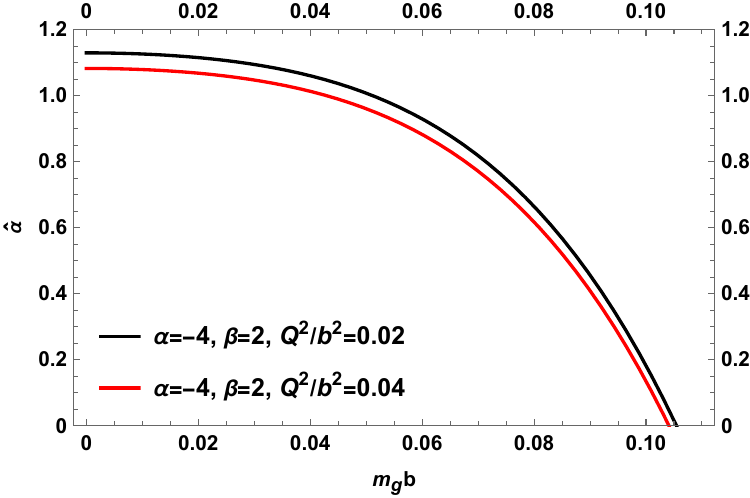}
\put(-100,-10){(d)}
\end{minipage}
\caption{Weak field limit deflection angle of black hole in dRGT massive gravity.
 (a) black curve represents Schwartzschild-AdS black hole and red curve represents RN-AdS black hole with $\gamma=0, \zeta=0$ (\,i.e. $\Lambda=-m_g^2<0$); (b) blue curve plotted with example values of $\alpha, \beta$ for $\Lambda<0, \gamma<0, \zeta>0$ and red curve plotted for $\Lambda<0, \gamma<0, \zeta<0$; (c) plotted with example values of $\alpha, \beta$ for $\Lambda=0, \gamma=0, \zeta>0$ and (d) plotted for $\Lambda>0, \gamma>0, \zeta>0$. Here we take $b=1, M/b=0.2, u_S=1/25, u_R=1/30$.}
{\label{alpha1Fig}}
\end{figure}

To check the effect of graviton mass parameter $m_g$, impact parameter $b$ on the deflection angle we likewise illustrate the deflection angle on plots in Fig.\ref{alpha1Fig}, in various possible situations.

Figure \ref{alpha1Fig} depicts the weak-field limit deflection angle $\hat{\alpha}$ of black holes in dRGT massive gravity (with $b=1$, $M/b=0.2$, $u_S=1/25$, $u_N=1/30$), focusing on its variation with the graviton mass $m_g$ (quantified by $m_g b$) across different configurations:
(a) For $\Lambda < 0$, $\gamma=0$, $\zeta=0$: Both the Schwarzschild-AdS (black, $\alpha=-1,\beta=1/3, Q =0$) and RN-AdS (red, $\alpha=-1,\beta=1/3,Q/b=0.2$) curves show $\hat{\alpha}$ decreasing sharply as $m_g b$ increases; the RN-AdS case (red) yields a slightly smaller $\hat{\alpha}$ than the Schwarzschild-AdS counterpart (black) at the same $m_g b$.
(b) For $\Lambda < 0$ (with $\gamma<0$, $\zeta\neq0$): With sample $\alpha=1/2,Q/ b=0.2$, $\hat{\alpha}$ declines with $m_g b$ (faster for the red curve ($\beta=-1/2$), $\zeta<0$, than the black curve ($\beta=1/2$), $\zeta>0$), and the initial $\hat{\alpha}$ (at small $m_g b$) is higher than in panel (a).
(c) For $\Lambda=0$, $\gamma=0$, $\zeta>0$: As $m_g b$ rises, $\hat{\alpha}$ drops gradually (slower than in (a) and (b)); increasing $Q^2/b^2$ (from $0.02$ (black) to $0.04$ (red), with $\alpha=-2,\beta=1$) reduces the initial $\hat{\alpha}$ but has a milder effect on its decreasing rate.
(d) For $\Lambda>0$, $\gamma>0$, $\zeta>0$: $\hat{\alpha}$ decreases slowly with $m_g b$ (slowest among all panels); a larger $Q^2/b^2$ ($0.04$ (red) vs. $0.02$ (black), with $\alpha=-4$, $\beta=2$) leads to a smaller $\hat{\alpha}$ at all $m_g b$, consistent with panel (c) but with a flatter overall trend.

In summary, $\hat{\alpha}$ generally decreases with increasing $m_g b$ across all cases, while the decreasing rate (and initial $\hat{\alpha}$) depends strongly on $\Lambda$, $\zeta$, and $Q^2/ b^2$ (e.g., faster decline for $\Lambda<0$, slower for $\Lambda>0$).

\section{The Effects of Graviton Mass on Einstein Rings For the Black Hole in dRGT Gravity}\label{observe}


In this section, we derive the analytical expressions of the angular radius of the Einstein ring for a new class of black hole in the framework of dRGT massive gravity. In gravitational lensing observations, physical observables are mostly constrained by lens equation, which is given by \cite{Bozza:2008ev,Izumi:2013tya,Jusufi:2018jof}
\begin{equation}\label{lens equation}
D_{RS}\tan{\ss}=\frac{D_{LR}\sin\Theta-D_{LS}\sin(\hat{\alpha}-\Theta)}{\cos(\hat{\alpha}-\Theta)}\,.
\end{equation}
Here, $D_{LR}$ is the distance between observer and lens plane, $D_{LS}$ is the distance between lens plane and source plane, $D_{RS}= D_{LR}+D_{LS}$ is the distance between observer and source plane. The angle ${\ss}$ denotes the angular position of particle source with respect to the ``optical axis'', and $\Theta$ is the angular position of the lensed image detected by observer. See corresponding schematic diagram of the weak gravitation lensing in literature \cite{Qiao:2021trw}. The images of the lensed objects can be achieved by solving lens equation (\ref{lens equation}). In the weak gravitational lensing, for distant source and observer, we have the approximations $\tan{\ss} \approx {\ss}$, $\sin\Theta \approx \Theta$, $\sin(\hat{\alpha}-\Theta) \approx \hat{\alpha}-\Theta$ and $\cos(\hat{\alpha}-\Theta) \approx 1$. Then the lens equation (\ref{lens equation}) reduces to \cite{Bozza:2001xd}
\begin{equation}\label{EinsteinRing}
 {\ss}=\Theta-\frac{D_{LS}}{D_{RS}}\hat{\alpha}\,.
\end{equation} 	
An Einstein ring is formed under the condition ${\ss} = 0$, which is the case for lens, receiver and source are perfectly aligned. And then, the above equation gives the angular radius of the Einstein ring as
\begin{eqnarray}\label{EinsteinRing1}
\Theta_{\text{E}} \simeq \frac{D_{LS}}{D_{RS}} \hat{\alpha} +\mathcal{O}(\varepsilon^3 ,...) \,.
\end{eqnarray}
Here $D_{LS}=1/u_S$ and $D_{LR}=1/u_R$.
Moreover, under this approximation, the impact parameter $b$ satisfies \cite{Gao:2023ltr}
\begin{equation}\label{approximationb}
b \approx D_{LR}\sin{\Theta_{\text{E}}}\simeq D_{LR}\Theta_{\text{E}} \,.
\end{equation}
Then, combining Eqs.(\ref{EinsteinRing1}) and (\ref{approximationb}), the angular radius of the Einstein ring $\Theta_{\text{E}}$ can be solved. But, it is difficult to get the roots of the solution without any assumptions. Therefore, for convenience of calculation, we consider three special cases with considerable assumptions in the following.

When we set $\Lambda=0$ and $\gamma=0$, the black hole becomes asymptotically flat. In this setup, we obtain the analytical form of the angular radius of Einstein ring for a new class of black hole in the framework of dRGT massive gravity given by
\begin{eqnarray}\label{ERAgm0}
\Theta_{\text{E}}\simeq \frac{\eta}{24 D_{LR}^2 D_{RS}} +\frac{\pi (\zeta -4) \zeta D_{LS}}{24 D_{RS}}-\frac{16 (\pi \zeta +\zeta -2) M D_{LR} D_{LS}}{\eta}+\frac{\pi^2 (\zeta-4)^2 \zeta ^2 D_{LR}^2 D_{LS}^2}{24 \eta D_{RS}} \,,
\end{eqnarray}
where
\begin{eqnarray}
\eta&=&\Big[\pi D_{LR}^4 D_{LS} \left(D_{LR}^2 \left(\pi ^2 (\zeta -4)^3 \zeta ^3 D_{LS}^2-K+20736 M^2-5184 Q^2\right)-K D_{LR} D_{LS}\right. \nonumber\\
&&\left.+10368 D_{LR} D_{LS} \left(4 M^2-Q^2\right)+5184 D_{LS}^2 \left(4 M^2-Q^2\right)\right) +24 \sqrt{6} \sqrt{\omega D_{LR}^8 D_{LS}^2 D_{RS}^2}\Big]^{1/3}
\end{eqnarray}
with
\begin{eqnarray}
\omega&=&64 D_{LR} D_{LS} \big[\delta +243 \pi^2 \left(Q^2-4 M^2\right)^2\big]+D_{LR}^2 \big[-\beta +64 \delta +7776 \pi ^2 \left(Q^2-4 M^2\right)^2\big]+7776 \pi ^2 D_{LS}^2 \left(Q^2-4 M^2\right)^2 \,,\nonumber\\
\delta&=&(\pi \zeta +\zeta -2) M D_{LS} \big[4 (64 (\zeta -4) \zeta +\pi [128(\zeta -2)+\pi (37 \zeta +108)] \zeta +256) M^2+27 \pi ^2 \zeta Q^2 (\zeta -4) \big]\,,\nonumber\\
\beta&=&\pi ^2 (\zeta -4)^2 \zeta ^2 D_{LS}^2 \big[4 (\zeta (\pi (16+5 \pi) \zeta +8 \zeta +4 \pi (3 \pi -8)-32)+32) M^2+3 \pi^2 (\zeta -4) \zeta Q^2\big]\,,\nonumber\\
K&=&576\zeta M D_{LS}[\zeta(\pi+1)-2](\zeta-4)  \,.\nonumber
\end{eqnarray}
In this case, besides the black hole's mass and electric charge, the parameter $\zeta$ also influences the light deflection angle. Notably, only $\zeta$ depends on the graviton mass--a key result indicating that the graviton mass exerts a non-trivial effect on stellar light deflection through this specific parameter. In the limit where graviton becomes massless, $i.e.,$ $m_g \rightarrow 0$, and neglecting higher-order terms, we obtain the approximate angular radius of the Einstein ring as
\begin{eqnarray}\label{ERAgmz0}
\Theta_{\text{E}}\simeq \frac{8 M D_{LR} D_{LS}}{3^{\frac{1}{3}}\xi^{\frac{1}{3}}}+\frac{{\xi}^\frac{1}{3}}{2\times 3^{2/3} D_{LR}^2 D_{RS}} \,,
\end{eqnarray}
where
\begin{eqnarray}
\xi =\sqrt{3} \sqrt{D_{LR}^8 D_{LS}^2 D_{RS}^3 (243 \pi^2 Q^4 D_{RS}-4096 M^3 D_{LR} D_{LS})}-27 \pi Q^2 D_{LR}^4 D_{LS} D_{RS}^2 \,.\,\,\,\,
\end{eqnarray}
This result is consistent with the angular radius of the Einstein ring for the Reissner-Nordstr\"{o}m (RN) black hole, confirming the expected recovery of the classical RN limit under the above approximations.
If we further set $Q=0$, we recover the angular radius of the Einstein ring for the Schwarzschild black hole, given by
\begin{eqnarray}\label{ERAgmzQ0}
\Theta_{\text{E}}\simeq \frac{2 \sqrt{M} \sqrt{D_{LS}}}{\sqrt{D_{LR} D_{RS}}} \,.
\end{eqnarray}

\section{Conclusion and Discussion}\label{summary}

In this paper, we investigate the gravitational lensing effect of a charged, spherically symmetric black hole (with nonzero graviton mass) in dRGT massive gravity via the Gauss-Bonnet theorem. Our calculations are carried out under the weak field approximation, where the dimensionless ratio $M/b$ (black hole mass parameter $M$ to impact parameter $b$) is treated as a first-order small quantity; additionally, we assume the graviton mass $m_g$ and charge $Q$ contribute only at the second-order small level. We systematically derive the light deflection angle up to the second-order approximation, with the light source and receiver positioned at finite distances from the black hole. When the impact parameter $b$ is much smaller than the source/receiver distance (with their ratio bounded by a first-order small quantity), the source/receiver can be approximated as infinitely distant--a valid simplification that allows us to further derive the analytical expression for the Einstein angular radius in the aligned (source-lens-receiver) configuration.

In the deflection angle analysis, we verify that setting $m_g = 0$ recovers the gravitational deflection of the conventional Reissner-Nordstr\"{o}m black hole. For nonzero $m_g$, however, the graviton mass introduces lensing effects distinctly different from those of RN black holes: our results confirm $m_g$ acts as a pivotal regulator of light deflection in dRGT massive gravity, inherently enhancing deflection--yet its specific influence (e.g., variation strength, saturation behavior, or even trend reversal) is not universal. These findings underscore the non-trivial role of $m_g$ in bridging dRGT's theoretical parameters to observable gravitational phenomena. We also note that the deflection angle $\hat\alpha$ decreases with increasing charge $Q$. Furthermore, we derive the analytical expression for the Einstein ring's angular radius in the framework of dRGT massive gravity, with the expression explicitly involving the graviton mass $m_g$. This result provides a foundational analytical basis for further investigating the dependence of the Einstein ring's properties on $m_g$ and comparing it with the corresponding quantity of traditional Reissner-Nordstr\"{o}m black holes. Such an explicit formulation may offer insights for future astrophysical studies aiming to probe modified gravity theories via weak lensing observations.

Notably, all results obtained in this paper have been compared with the cases of conventional Reissner-Nordstr\"{o}m black holes ($m_g\rightarrow 0$), Reissner-Nordstr\"{o}m-AdS black holes ($\gamma\rightarrow 0, \zeta\rightarrow 0$), and Schwarzschild black holes ($m_g\rightarrow 0, Q\rightarrow 0$). We hope that the conclusions of this paper can provide useful references for theoretical and experimental/observational studies of black holes with nonzero graviton masses. At the same time, we also hope that the work presented in this paper can offer some new insights into black hole physics and analogous gravitational models in high-energy physics and condensed matter physics. Looking ahead, we plan to further explore more properties of black holes within the framework of dRGT massive gravity and search for possible observational evidence to validate these theoretical predictions. Current observational constraints on graviton mass from 43 GWTC-3 events $m_g \lesssim 2.42\times10^{-23}\,\text{eV}/c^2$\cite{LIGOScientific:2021sio} and Solar System bound $m_g \lesssim 3.16\times10^{-23}\,\text{eV}/c^2$ \cite{Bernus:2020szc} provide critical boundaries for our model; incorporating these joint constraints into our lensing framework will be the focus of future work.


\section{Acknowledgments}

The authors thank the referees for their valuable comments, which have greatly improved this paper. We also thank Dr. X. J. Gao for helpful discussions. Haximjan Abdusattar is supported by Kashi University high-level talent research start-up fund project (Grant No. 022024002), and Tianchi Talented Young Doctors Program of Xinjiang Uyghur Autonomous Region.
Shi-Bei Kong is supported by the National Natural Science Foundation of China (Grant No. 12465011) and East China University of Technology (Grant No. DHBK2023002).


{}


\begin{thebibliography}{10}

\bibitem{Phillips:1993ng}
M.~M.~Phillips,
``The absolute magnitudes of Type IA supernovae'',
{\hypersetup{urlcolor=vividviolet}\href{https://inspirehep.net/literature/791426}{Astrophys. J. Lett. \textbf{413} (1993), L105-L108}}.

\bibitem{SupernovaSearchTeam:1998fmf}
A.~G.~Riess \textit{et al.} [Supernova Search Team],
``Observational evidence from supernovae for an accelerating universe and a cosmological constant'',
{\hypersetup{urlcolor=vividviolet}\href{https://inspirehep.net/literature/470671}{Astron. J. \textbf{116} (1998), 1009-1038}}, \href{https://arxiv.org/abs/astro-ph/9805201}{[arXiv:astro-ph/9805201]}.

\bibitem{Planck:2015fie}
P.~A.~R.~Ade \textit{et al.} [Planck],
``Planck 2015 results. XIII. Cosmological parameters'',
{\hypersetup{urlcolor=vividviolet}\href{https://inspirehep.net/literature/1343079}{Astron. Astrophys. \textbf{594} (2016), A13}}, \href{https://arxiv.org/abs/1502.01589}{[arXiv:astro-ph.CO/1502.01589]}.

\bibitem{WMAP:2003elm}
D.~N.~Spergel \textit{et al.} [WMAP],
``First year Wilkinson Microwave Anisotropy Probe (WMAP) observations: Determination of cosmological parameters'',
{\hypersetup{urlcolor=vividviolet}\href{https://inspirehep.net/literature/613135}{Astrophys. J. Suppl. \textbf{148} (2003), 175-194}}, \href{https://arxiv.org/abs/astro-ph/0302209}{[arXiv:astro-ph/0302209]}.


\bibitem{Beutler:2011hx}
F.~Beutler, C.~Blake, M.~Colless, D.~H.~Jones, L.~Staveley-Smith, L.~Campbell, Q.~Parker, W.~Saunders and F.~Watson,
``The $6$dF Galaxy Survey: Baryon Acoustic Oscillations and the Local Hubble Constant'',
{\hypersetup{urlcolor=vividviolet}\href{https://inspirehep.net/literature/914162}{Mon. Not. Roy. Astron. Soc. \textbf{416} (2011), 3017-3032}}, \href{https://arxiv.org/abs/1106.3366}{[arXiv:astro-ph.CO/1106.3366]}.

\bibitem{SDSS:2009ocz}
W.~J.~Percival \textit{et al.} [SDSS],
``Baryon Acoustic Oscillations in the Sloan Digital Sky Survey Data Release $7$ Galaxy Sample'',
{\hypersetup{urlcolor=vividviolet}\href{https://inspirehep.net/literature/825356}{Mon. Not. Roy. Astron. Soc. \textbf{401} (2010), 2148-2168}}, \href{https://arxiv.org/abs/0907.1660}{[arXiv:astro-ph.CO/0907.1660]}.


\bibitem{Weinberg:1988cp}
S.~Weinberg,
``The Cosmological Constant Problem'',
{\hypersetup{urlcolor=vividviolet}\href{https://inspirehep.net/literature/263386}{Rev. Mod. Phys. \textbf{61} (1989), 1-23}}.

\bibitem{Peebles:2002gy}
P.~J.~E.~Peebles and B.~Ratra,
``The Cosmological Constant and Dark Energy'',
{\hypersetup{urlcolor=vividviolet}\href{https://inspirehep.net/literature/590673}{Rev. Mod. Phys. \textbf{75} (2003), 559-606}}, \href{https://arxiv.org/abs/astro-ph/0207347}{[arXiv:astro-ph/0207347]}.

\bibitem{deRham:2010ik}
C.~de Rham and G.~Gabadadze,
``Generalization of the Fierz-Pauli Action'',
{\hypersetup{urlcolor=vividviolet}\href{https://inspirehep.net/literature/860500}{Phys. Rev. D \textbf{82} (2010), 044020}}, \href{https://arxiv.org/abs/1007.0443}{[arXiv:hep-th/1007.0443]}.

\bibitem{deRham:2014zqa}
C.~de Rham,
``Massive Gravity'',
{\hypersetup{urlcolor=vividviolet}\href{https://inspirehep.net/literature/1278081}{Living Rev. Rel. \textbf{17} (2014), 7}}, \href{https://arxiv.org/abs/1401.4173}{[arXiv:hep-th/1401.4173]}.

\bibitem{Psaltis:2008bb}
D.~Psaltis,
``Probes and Tests of Strong-Field Gravity with Observations in the Electromagnetic Spectrum'',
{\hypersetup{urlcolor=vividviolet}\href{https://inspirehep.net/literature/787729}{Living Rev. Rel. \textbf{11} (2008), 9}}, \href{https://arxiv.org/abs/0806.1531}{[arXiv:astro-ph/0806.1531]}.

\bibitem{Cai:2014znn}
R.~G.~Cai, Y.~P.~Hu, Q.~Y.~Pan and Y.~L.~Zhang,
``Thermodynamics of Black Holes in Massive Gravity'',
{\hypersetup{urlcolor=vividviolet}\href{https://inspirehep.net/literature/1315428}{Phys. Rev. D \textbf{91} (2015) no.2, 024032}}, \href{https://arxiv.org/abs/1409.2369}{[arXiv:hep-th/1409.2369]}.

\bibitem{Xu:2015rfa}
J.~Xu, L.~M.~Cao and Y.~P.~Hu,
``$P$-$V$ criticality in the extended phase space of black holes in massive gravity'',
{\hypersetup{urlcolor=vividviolet}\href{https://inspirehep.net/literature/1375807}{Phys. Rev. D \textbf{91} (2015) no.12, 124033}}, \href{https://arxiv.org/abs/1506.03578}{[arXiv:gr-qc/1506.03578]}.


\bibitem{Ghosh:2015cva}
S.~G.~Ghosh, L.~Tannukij and P.~Wongjun,
``A class of black holes in dRGT massive gravity and their thermodynamical properties'',
{\hypersetup{urlcolor=vividviolet}\href{https://inspirehep.net/literature/1377560}{Eur. Phys. J. C \textbf{76} (2016) no.3, 119}}, \href{https://arxiv.org/abs/1506.07119}{[arXiv:gr-qc/1506.07119]}.


\bibitem{Chabab:2019mlu}
M.~Chabab, H.~El Moumni, S.~Iraoui and K.~Masmar,
``Phase transitions and geothermodynamics of black holes in dRGT massive gravity'',
{\hypersetup{urlcolor=vividviolet}\href{https://inspirehep.net/literature/1728696}{Eur. Phys. J. C \textbf{79} (2019) no.4, 342}}, \href{https://arxiv.org/abs/1904.03532}{[arXiv:hep-th/1904.03532]}.


\bibitem{Hou:2020yni}
M.~S.~Hou, H.~Xu and Y.~C.~Ong,
``Hawking Evaporation of Black Holes in Massive Gravity'',
{\hypersetup{urlcolor=vividviolet}\href{https://inspirehep.net/literature/1813012}{Eur. Phys. J. C \textbf{80} (2020) no.11, 1090}}, \href{https://arxiv.org/abs/2008.10049}{[arXiv:hep-th/2008.10049]}.

\bibitem{Nam:2020gud}
C.~H.~Nam,
``Effect of massive gravity on Joule\textendash{}Thomson expansion of the charged AdS black hole'',
{\hypersetup{urlcolor=vividviolet}\href{https://inspirehep.net/literature/1781899}{Eur. Phys. J. Plus \textbf{135} (2020) no.2, 259}}.


\bibitem{Bartelmann:1999yn}
M.~Bartelmann and P.~Schneider,
``Weak gravitational lensing",
{\hypersetup{urlcolor=vividviolet}\href{https://inspirehep.net/literature/521013}{Phys. Rept. \textbf{340} (2001), 291-472}}, \href{https://arxiv.org/abs/astro-ph/9912508}{[arXiv:astro-ph/9912508]}.

\bibitem{Schmidt:2008hc}
F.~Schmidt,
``Weak Lensing Probes of Modified Gravity'',
{\hypersetup{urlcolor=vividviolet}\href{https://inspirehep.net/literature/787445}{Phys. Rev. D \textbf{78} (2008), 043002}}, \href{https://arxiv.org/abs/0805.4812}{[arXiv:astro-ph/0805.4812]}.


\bibitem{Hoekstra:2013via}
H.~Hoekstra, M.~Bartelmann, H.~Dahle, H.~Israel, M.~Limousin and M.~Meneghetti,
``Masses of galaxy clusters from gravitational lensing'',
{\hypersetup{urlcolor=vividviolet}\href{https://inspirehep.net/literature/1223789}{Space Sci. Rev. \textbf{177} (2013), 75-118}}, \href{https://arxiv.org/abs/1303.3274}{[arXiv:astro-ph.CO/1303.3274]}.


\bibitem{Gibbons:2008rj}
 G. W. Gibbons and M. C. Werner,
 ``Applications of the Gauss-Bonnet theorem to gravitational lensing",
 {\hypersetup{urlcolor=vividviolet}\href{https://inspirehep.net/literature/790009}{Class. Quant. Grav. \textbf{25} (2008), 235009}}, \href{https://arxiv.org/abs/0807.0854}{[arXiv:gr-qc/0807.0854]}.
	
\bibitem{Ishihara:2016vdc}
A.~Ishihara, Y.~Suzuki, T.~Ono, T.~Kitamura and H.~Asada,
``Gravitational bending angle of light for finite distance and the Gauss-Bonnet theorem'',
{\hypersetup{urlcolor=vividviolet}\href{https://inspirehep.net/literature/1452798}{Phys. Rev. D \textbf{94} (2016) no.8, 084015}}, \href{https://arxiv.org/abs/1604.08308}{[arXiv:gr-qc/1604.08308]}.
	
\bibitem{Ishihara:2016sfv}
A.~Ishihara, Y.~Suzuki, T.~Ono and H.~Asada,
``Finite-distance corrections to the gravitational bending angle of light in the strong deflection limit'',
{\hypersetup{urlcolor=vividviolet}\href{https://inspirehep.net/literature/1503132}{Phys. Rev. D \textbf{95} (2017) no.4, 044017}}, \href{https://arxiv.org/abs/1612.04044}{[arXiv:gr-qc/1612.04044]}.

%

\bibitem{Ovgun:2018tua}		
A. \"Ovg\"un, I. Sakalli and J. Saavedra,
``Shadow cast and deflection angle of Kerr-Newman-Kasuya spacetime", {\hypersetup{urlcolor=vividviolet}\href{https://inspirehep.net/literature/1680506}{JCAP \textbf{10} (2018), 041}}, \href{https://arxiv.org/abs/1807.00388}{arXiv:gr-qc/1807.00388]}.

\bibitem{Ono:2019hkw}
T.~Ono and H.~Asada,
``The effects of finite distance on the gravitational deflection angle of light'',
{\hypersetup{urlcolor=vividviolet}\href{https://inspirehep.net/literature/1738689}{Universe \textbf{5} (2019) no.11, 218}}, \href{https://arxiv.org/abs/1906.02414}{[arXiv:gr-qc/1906.02414]}.

\bibitem{Jusufi:2018jof}
K.~Jusufi, A.~\"Ovg\"un, J.~Saavedra, Y.~V\'asquez and P.~A.~Gonz\'alez,
``Deflection of light by rotating regular black holes using the Gauss-Bonnet theorem'',
{\hypersetup{urlcolor=vividviolet}\href{https://inspirehep.net/literature/1665590}{Phys. Rev. D \textbf{97} (2018) no.12, 124024}}, \href{https://arxiv.org/abs/1804.00643}{[arXiv:gr-qc/1804.00643]};
K. Jusufi,
Determining the Topology and Deflection Angle of Ringholes via Gauss-Bonnet Theorem, {\hypersetup{urlcolor=vividviolet}\href{https://doi.org/10.3390/universe7020044}{Universe {\bf 7}, 44 (2021)}}, \href{arXiv:1807.09748}{[arXiv:gr-qc/1807.09748]}.

\bibitem{Crisnejo:2018uyn}	
G. Crisnejo and E. Gallo,
``Weak lensing in a plasma medium and gravitational deflection of massive particles using the Gauss-Bonnet theorem. A unified treatment",
{\hypersetup{urlcolor=vividviolet}\href{https://doi.org/10.1103/PhysRevD.97.124016}{Phys. Rev. D \textbf{97} (2018) no.12, 124016}}, \href{https://arxiv.org/abs/1804.05473}{[arXiv:gr-qc/1804.05473]};	
G. Crisnejo, E. Gallo and A. Rogers,
``Finite distance corrections to the light deflection in a gravitational field with a plasma medium", {\hypersetup{urlcolor=vividviolet}\href{https://doi.org/10.1103/PhysRevD.99.124001}{Phys. Rev. D {\bf 99}, 124001 (2019)}}, \href{https://arxiv.org/abs/1807.00724}{[arXiv:gr-qc/1807.00724]}.


\bibitem{Pantig:2020odu}		
R. C. Pantig and E. T. Rodulfo,
``Weak deflection angle of a dirty black hole", {\hypersetup{urlcolor=vividviolet}\href{https://doi.org/10.1016/j.cjph.2020.06.015}{Chin. J. Phys. \textbf{66} (2020), 691-702}}, \href{https://arxiv.org/abs/2003.00764}{[arXiv:gr-qc/2003.00764]}.
	
\bibitem{Fu:2021akc}
Qi-Ming Fu, Li Zhao and Yu-Xiao Liu,
``Weak deflection angle by electrically and magnetically charged black holes from nonlinear electrodynamics", {\hypersetup{urlcolor=vividviolet}\href{https://doi.org/10.1103/PhysRevD.104.024033}{Phys. Rev. D {\bf 104}, 024033 (2021)}}, \href{https://arxiv.org/abs/2101.08409}{[arXiv:gr-qc/2101.08409]}.
	

\bibitem{Javed:2020fli}
W.~Javed, M.~B.~Khadim, A.~\"Ovg\"un and J.~Abbas,
``Weak gravitational lensing by stringy black holes'',
{\hypersetup{urlcolor=vividviolet}\href{https://inspirehep.net/literature/1787897}{Eur. Phys. J. Plus \textbf{135} (2020) no.3, 314}}, \href{https://arxiv.org/abs/2004.00408}{[arXiv:gr-qc/2004.00408]};
W.~Javed, M.~B.~Khadim and A.~\"Ovg\"un,
``Weak gravitational lensing by Bocharova\textendash{}Bronnikov\textendash{}Melnikov\textendash{}Bekenstein black holes using Gauss\textendash{}Bonnet theorem'',
{\hypersetup{urlcolor=vividviolet}\href{https://inspirehep.net/literature/1808787}{Eur. Phys. J. Plus \textbf{135} (2020) no.7, 595}}, \href{https://arxiv.org/abs/2007.14844}{[arXiv:gr-qc/2007.14844]};
W.~Javed, J.~Abbas and A.~\"Ovg\"un,
``Deflection angle of photon from magnetized black hole and effect of nonlinear electrodynamics'',
{\hypersetup{urlcolor=vividviolet}\href{https://inspirehep.net/literature/1750702}{Eur. Phys. J. C \textbf{79} (2019) no.8, 694}}, \href{https://arxiv.org/abs/1908.09632}{[arXiv:physics.gen-ph/1908.09632]}.
	


\bibitem{Ovgun:2018fnk}
A.~\"Ovg\"un,
``Light deflection by Damour-Solodukhin wormholes and Gauss-Bonnet theorem'',
{\hypersetup{urlcolor=vividviolet}\href{https://inspirehep.net/literature/1673418}{Phys. Rev. D \textbf{98} (2018) no.4, 044033}}, \href{https://arxiv.org/abs/1805.06296}{[arXiv:gr-qc/1805.06296]}.

\bibitem{Jusufi:2017uhh}
K.~Jusufi and A.~Ovg\"un,
``Effect of the cosmological constant on the deflection angle by a rotating cosmic string'',
{\hypersetup{urlcolor=vividviolet}\href{https://inspirehep.net/literature/1641067}{Phys. Rev. D \textbf{97} (2018) no.6, 064030}}, \href{https://arxiv.org/abs/1712.01771}{[arXiv:gr-qc/1712.01771]};
K.~Jusufi, M.~C.~Werner, A.~Banerjee and A.~\"Ovg\"un,
{\hypersetup{urlcolor=vividviolet}\href{https://inspirehep.net/literature/1514228}{Phys. Rev. D \textbf{95} (2017) no.10, 104012}}, \href{https://arxiv.org/abs/1702.05600}{[arXiv:gr-qc/1702.05600]}.

\bibitem{deLeon:2019qnp}
K.~de Leon and I.~Vega,
``Weak gravitational deflection by two-power-law densities using the Gauss-Bonnet theorem'',
{\hypersetup{urlcolor=vividviolet}\href{https://inspirehep.net/literature/1725499}{Phys. Rev. D \textbf{99} (2019) no.12, 124007}}, \href{https://arxiv.org/abs/1903.06951}{[arXiv:gr-qc/1903.06951]}.

\bibitem{Ono:2018jrv}
T.~Ono, A.~Ishihara and H.~Asada,
``Deflection angle of light for an observer and source at finite distance from a rotating global monopole'',
{\hypersetup{urlcolor=vividviolet}\href{https://inspirehep.net/literature/1702227}{Phys. Rev. D \textbf{99} (2019) no.12, 124030}},  \href{https://arxiv.org/abs/1811.01739}{[arXiv:gr-qc/1811.01739]}.


\bibitem{Crisnejo:2019ril}	
G. Crisnejo, E. Gallo and K. Jusufi,
``Higher order corrections to deflection angle of massive particles and light rays in plasma media for stationary spacetimes using the Gauss-Bonnet theorem", {\hypersetup{urlcolor=vividviolet}\href{https://doi.org/10.1103/PhysRevD.100.104045}{Phys. Rev. D \textbf{100} (2019) no.10, 104045}}, \href{https://arxiv.org/abs/1910.02030}{[arXiv:gr-qc/1910.02030]}.
	
\bibitem{Ovgun:2019wej}
A.~\"Ovg\"un,
``Weak field deflection angle by regular black holes with cosmic strings using the Gauss-Bonnet theorem'',
{\hypersetup{urlcolor=vividviolet}\href{https://inspirehep.net/literature/1720013}{Phys. Rev. D \textbf{99} (2019) no.10, 104075}},  \href{https://arxiv.org/abs/1902.04411}{[arXiv:gr-qc/1902.04411]}.


\bibitem{Li:2020zxi}
Z.~Li and T.~Zhou,
``Kerr black hole surrounded by a cloud of strings and its weak gravitational lensing in Rastall gravity'',
{\hypersetup{urlcolor=vividviolet}\href{https://inspirehep.net/literature/1774157}{Phys. Rev. D \textbf{104} (2021) no.10, 104044}}, \href{https://arxiv.org/abs/2001.01642}{[arXiv:gr-qc/2001.01642]};
Z. -H. Li, G. -S. He and T. Zhou,
``Gravitational deflection of relativistic massive particles by wormholes", {\hypersetup{urlcolor=vividviolet}\href{https://doi.org/10.1103/PhysRevD.101.044001}{Phys. Rev. D {\bf 101}, 044001 (2020)}}, \href{https://arxiv.org/abs/1908.01647}{[arXiv:gr-qc/1908.01647]};
Z. -H. Li, J. -J. Jia,
``The finite-distance gravitational deflection of massive particles in stationary spacetime: a Jacobi metric approach", {\hypersetup{urlcolor=vividviolet}\href{https://doi.org/10.1140/epjc/s10052-020-7665-8}{Eur. Phys. J. C {\bf 80}, 157 (2020)}}, \href{https://arxiv.org/abs/1912.05194}{[arXiv:gr-qc/1912.05194]}.
	

\bibitem{He:2020eah}
G.~He, X.~Zhou, Z.~Feng, X.~Mu, H.~Wang, W.~Li, C.~Pan and W.~Lin,
``Gravitational deflection of massive particles in Schwarzschild-de Sitter spacetime'',
\href{https://inspirehep.net/literature/1816224}{Eur. Phys. J. C \textbf{80} (2020) no.9, 835}.

\bibitem{Pantig:2022toh}
R.~C.~Pantig and A.~\"Ovg\"un,
``Dark matter effect on the weak deflection angle by black holes at the center of Milky Way and M87 galaxies'',
{\hypersetup{urlcolor=vividviolet}\href{https://inspirehep.net/literature/2007036}{Gen. Rel. Grav. \textbf{50} (2018) no.5, 48}}, \href{https://arxiv.org/abs/2201.03365}{[arXiv:gr-qc/2201.03365]}.

\bibitem{Qiao:2021trw}
C.~K.~Qiao and M.~Zhou,
``The gravitational bending of acoustic Schwarzschild black hole'',
{\hypersetup{urlcolor=vividviolet}\href{https://inspirehep.net/literature/1921029}{Eur. Phys. J. C \textbf{83}, no.4, 271 (2023)} }, \href{https://arxiv.org/abs/2109.05828}{[arXiv:gr-qc/2109.05828]}.


\bibitem{Arakida:2017hrm}
H.~Arakida,
``Light deflection and Gauss-Bonnet theorem: definition of total deflection angle and its applications'',
{\hypersetup{urlcolor=vividviolet}\href{https://inspirehep.net/literature/1615860}{Gen. Rel. Grav. \textbf{50} (2018) no.5, 48}}, \href{https://arxiv.org/abs/1708.04011}{[arXiv:gr-qc/1708.04011]}.


\bibitem{Takizawa:2020egm}
K. Takizawa, T. Ono, and H. Asada,
``Gravitational deflection angle of light: Definition by an observer and its application to an asymptotically nonflat spacetime",
{\hypersetup{urlcolor=vividviolet}\href{https://doi.org/10.1103/PhysRevD.101.104032}{Phys. Rev. D \textbf{101} (2020) no.10, 104032}}, \href{https://arxiv.org/abs/2001.03290}{[arXiv:gr-qc/2001.03290]}.
	
\bibitem{Haroon:2018ryd}
S.~Haroon, M.~Jamil, K.~Jusufi, K.~Lin and R.~B.~Mann,
``Shadow and Deflection Angle of Rotating Black Holes in Perfect Fluid Dark Matter with a Cosmological Constant'',
{\hypersetup{urlcolor=vividviolet}\href{https://inspirehep.net/literature/1697538}{Phys. Rev. D \textbf{99} (2019) no.4, 044015}}, \href{https://arxiv.org/abs/1810.04103}{[arXiv:gr-qc/1810.04103]}.

\bibitem{Atamurotov:2021hoq}
F.~Atamurotov, A.~Abdujabbarov and W.~B.~Han,
``Effect of plasma on gravitational lensing by a Schwarzschild black hole immersed in perfect fluid dark matter'',
\href{https://inspirehep.net/literature/1936182}{Phys. Rev. D \textbf{104} (2021) no.8, 084015};
F.~Atamurotov, U.~Papnoi and K.~Jusufi,
``Shadow and deflection angle of charged rotating black hole surrounded by perfect fluid dark matter'',
{\hypersetup{urlcolor=vividviolet}\href{https://inspirehep.net/literature/1861583}{Class. Quant. Grav. \textbf{39} (2022) no.2, 025014}},\href{https://arxiv.org/abs/2104.14898}{[arXiv:gr-qc/2104.14898]}.
	
\bibitem{Gao:2023ltr}
X.~J.~Gao, X.~k.~Yan, Y.~Yin and Y.~P.~Hu,
``Gravitational lensing by a charged spherically symmetric black hole immersed in thin dark matter,''
{\hypersetup{urlcolor=vividviolet}\href{https://inspirehep.net/literature/1861583}{Eur. Phys. J. C \textbf{83}, no.4, 281 (2023)}},\href{https://arxiv.org/abs/2311.11780}{[arXiv:gr-qc/2303.00190]}.

\bibitem{Panpanich:2019mll}
S.~Panpanich, S.~Ponglertsakul and L.~Tannukij,
``Particle motions and Gravitational Lensing in de Rham-Gabadadze-Tolley Massive Gravity Theory'',
{\hypersetup{urlcolor=vividviolet}\href{https://inspirehep.net/literature/1728518}{Phys. Rev. D \textbf{100} (2019) no.4, 044031}}, \href{https://arxiv.org/abs/1904.02915}{[arXiv:gr-qc/1904.02915]}.


\bibitem{Bozza:2002zj}
V.~Bozza,
``Gravitational lensing in the strong field limit'',
{\hypersetup{urlcolor=vividviolet}\href{https://inspirehep.net/literature/593539}{Phys. Rev. D \textbf{66} (2002), 103001} }, \href{https://arxiv.org/abs/gr-qc/0208075}{[arXiv:gr-qc/0208075]}.


\bibitem{Gibbons:2008hb}
G. W. Gibbons and C. M. Warnick,
``Universal properties of the near-horizon optical geometry", {\hypersetup{urlcolor=vividviolet}\href{https://inspirehep.net/literature/796055}{Phys. Rev. D \textbf{79} (2009), 064031}}, \href{https://arxiv.org/abs/0809.1571}{[arXiv:gr-qc/0809.1571]}.

\bibitem{Li:2019vhp}
Z.~Li, G.~He and T.~Zhou,
Gravitational deflection of relativistic massive particles by wormholes,
Phys. Rev. D \textbf{101}, no.4, 044001 (2020).

\bibitem{Claudel:2000yi}
C.~M.~Claudel, K.~S.~Virbhadra and G.~F.~R.~Ellis,
``The Geometry of photon surfaces'',
J. Math. Phys. \textbf{42}, 818-838 (2001)
[arXiv:gr-qc/0005050].

\bibitem{Eiroa:2002mk}
E.~F.~Eiroa, G.~E.~Romero and D.~F.~Torres,
Reissner-Nordstrom black hole lensing,
Phys. Rev. D \textbf{66}, 024010 (2002).

\bibitem{Crisnejo:2019xtp}
G.~Crisnejo, E.~Gallo and J.~R.~Villanueva,
``Gravitational lensing in dispersive media and deflection angle of charged massive particles in terms of curvature scalars and energy-momentum tensor'',
{\hypersetup{urlcolor=vividviolet}\href{https://inspirehep.net/literature/1733290}{Phys. Rev. D \textbf{100}, no.4, 044006 (2019)}}, \href{https://arxiv.org/abs/1905.02125}{[arXiv:gr-qc/1905.02125]}.


\bibitem{Bozza:2008ev}
V.~Bozza,
``A Comparison of approximate gravitational lens equations and a proposal for an improved new one'',
{\hypersetup{urlcolor=vividviolet}\href{https://inspirehep.net/literature/791426}{Phys. Rev. D \textbf{78} (2008), 103005}},\href{https://arxiv.org/abs/0807.3872}{[arXiv:gr-qc/0807.3872]}.


\bibitem{Izumi:2013tya}
K.~Izumi, C.~Hagiwara, K.~Nakajima, T.~Kitamura and H.~Asada,
``Gravitational lensing shear by an exotic lens object with negative convergence or negative mass'',
{\hypersetup{urlcolor=vividviolet}\href{https://inspirehep.net/literature/1235019}{Phys. Rev. D \textbf{88} (2013), 024049}},\href{https://arxiv.org/abs/1305.5037}{[arXiv:gr-qc/1305.5037]}.

\bibitem{Bozza:2001xd}
V.~Bozza, S.~Capozziello, G.~Iovane and G.~Scarpetta,
``Strong field limit of black hole gravitational lensing'',
{\hypersetup{urlcolor=vividviolet}\href{https://inspirehep.net/literature/553085}{Gen. Rel. Grav. \textbf{33} (2001), 1535-1548}}, \href{https://arxiv.org/abs/gr-qc/0102068}{[arXiv:gr-qc/0102068]}.


\bibitem{LIGOScientific:2021sio}
R.~Abbott \textit{et al.} [LIGO Scientific, VIRGO and KAGRA],
``Tests of General Relativity with GWTC-3,''
Phys. Rev. D \textbf{112}, no.8, 084080 (2025)
[arXiv:gr-qc/2112.06861].

\bibitem{Bernus:2020szc}
L.~Bernus, O.~Minazzoli, A.~Fienga, M.~Gastineau, J.~Laskar, P.~Deram and A.~Di Ruscio,
Phys. Rev. D \textbf{102}, no.2, 021501 (2020)
[arXiv:gr-qc/2006.12304].


\end{thebibliography}
\end{document}